\documentclass[showpacs,pra,twocolumn,aps]{revtex4}
 \usepackage{bm}
 \usepackage{amsmath}
 \usepackage{amssymb}
 \usepackage{latexsym} 
 \usepackage{amsfonts} 
 \usepackage{epsfig}
 \usepackage{psfrag} 

 \newcommand{\kf}{{\boldsymbol k}}

 \newcommand{\nn}{\nonumber\\} 
  
 \newcommand{\f}[1]{\mbox{\boldmath$#1$}}

 \newcommand{\na}{\mbox{\boldmath$\nabla$}}
 \newcommand{\bea}{\begin{eqnarray}}
 \newcommand{\ea}{\end{eqnarray}}
 \newcommand{\eea}{\end{eqnarray}}
 \newcommand{\ord}{{\cal O}}

\begin{document}

\title{Mean-field expansion in Bose-Einstein condensates
  with finite-range interactions}

\author{Ralf Sch\"utzhold, Michael Uhlmann, and Yan Xu}

\affiliation{Institut f\"ur Theoretische Physik, 
Technische Universit\"at Dresden, D-01062 Dresden, Germany}

\author{Uwe R.~Fischer}

\affiliation{Eberhard-Karls-Universit\"at T\"ubingen, 
Institut f\"ur Theoretische Physik,\\  
Auf der Morgenstelle 14, D-72076 T\"ubingen, Germany}

\begin{abstract} 
We present a formal derivation of the mean-field expansion for dilute
Bose-Einstein condensates with two-particle interaction potentials
which are weak and finite-range, but otherwise arbitrary.
The expansion allows for a controlled investigation of the impact of
microscopic interaction details (e.g., the scaling behavior) %###
on the mean-field approach and the induced higher-order corrections
beyond the $s$-wave scattering approximation. 
%###
%As an example, we calculate the quantum depletion in the presence of
%dipole-dipole interactions (in addition to the usual contact potential).
%###
\end{abstract}

  
\maketitle

%%%%%%%%%%%%%%%%%%%%%%%%%%%%%%%%%%%%%%%%%%%%%%%%%%%%%%%%%%%%%%%%%%%%%%%%%%%%%%%
%%%%%%%%%%%%%%%%%%%%%%%%%%%%%%%%%%%%%%%%%%%%%%%%%%%%%%%%%%%%%%%%%%%%%%%%%%%%%%%
\section{Introduction}
%%%%%%%%%%%%%%%%%%%%%%%%%%%%%%%%%%%%%%%%%%%%%%%%%%%%%%%%%%%%%%%%%%%%%%%%%%%%%%%
%%%%%%%%%%%%%%%%%%%%%%%%%%%%%%%%%%%%%%%%%%%%%%%%%%%%%%%%%%%%%%%%%%%%%%%%%%%%%%%

The experimental realization of Bose-Einstein condensation in 
atomic or molecular gases has rekindled interest in the quantum theory
of dilute bosonic gases \cite{Bogoliubov,Lee,Beliaev,Andersen}. 
So far, most investigations were based on the $s$-wave
scattering approximation which neglects microscopic details of
the two-particle interaction 
(such as attractive regions of the potential), 
and replaces the true interaction by a contact interaction
pseudopotential. 
To describe the impact of a possibly complicated microscopic
structure of the potential on quantities like the occupation number  
of the condensate, it is however necessary to go beyond the $s$-wave
approximation.  
On the other hand, the current theoretical description of
Bose-Einstein condensates is predominantly based on a mean-field
expansion around a macroscopically occupied state representing the
condensate.  In leading order this expansion 
yields the Gross-Pitaevski\v\i\/ equation, which forms the basis of 
most theoretical approaches to Bose-Einstein condensed gases 
\cite{BECreview}.  

The scope of the present article is to investigate consequences of 
deviations from pure $s$-wave scattering for the applicability of this   
mean-field expansion, and to derive the general zero-temperature 
equations governing the condensate and quantum fluctuations above it.
To this end, we present an {\it ab initio} approach,
controlled by properties of the bare interaction potential in the  
microscopic second-quantized equation of motion. 
While the presented approach is essentially restricted to sufficiently 
``weak'', that is, integrable potentials, we do %on the other hand 
not use pseudopotential or T-matrix approaches 
\cite{Griffin,Rusch,Olshanii}, 
for which the direct connection to microscopic properties 
of the interaction potential is lost. 
Our approach can thus be viewed as
being complementary to the latter treatments. In particular, if one 
can separate the potential into a strong ultra-short-range part, 
assumed to be replaceable by an integrable (pseudo-)potential, 
and a weaker remaining contribution with longer range, our analysis
%### longer 
applies to this latter part as a low-energy, effective description.
 
We test the validity  of the mean-field expansion, which is usually an 
{\em ad hoc} assumption, employing well-defined scaling properties 
of the interaction potential with the number of particles and thus the 
density of the system, under the prescription that the system always
remains dilute.  
It turns out that the consistency of the mean-field expansion
sensitively depends on the scaling exponent of the interaction
potential, and thus on the (formal) dependence of the interaction
range on the particle number. 
%###
This formal dependence on the particle number can be made more
intuitive by comparing the dilute-gas limit with the thermodynamic
limit. 

The structure of the paper is as follows. In the next Section we derive, 
using a number-conserving mean-field expansion, the equations of motion 
for the mean field itself (a modified Gross-Pitaevski\v\i\/ equation) and for 
the single- and multi-particle excitations above mean-field. In the third
Section, the consistency of this mean-field expansion is tested with three
scaling behaviors of the particle interaction potential. The 
fourth Section treats, as an explicit example, a dipole-dipole interaction 
potential. Even though such a potential is on the borderline of applicability
of our approach [because it is (just) not integrable in three spatial 
dimensions], we show that the 
calculation of certain quantities like the quantum depletion of the
condensate is still feasible.

%%%%%%%%%%%%%%%%%%%%%%%%%%%%%%%%%%%%%%%%%%%%%%%%%%%%%%%%%%%%%%%%%%%%%%%%%%%%%%%
%%%%%%%%%%%%%%%%%%%%%%%%%%%%%%%%%%%%%%%%%%%%%%%%%%%%%%%%%%%%%%%%%%%%%%%%%%%%%%%
\section{Number-conserving mean-field expansion} 
%%%%%%%%%%%%%%%%%%%%%%%%%%%%%%%%%%%%%%%%%%%%%%%%%%%%%%%%%%%%%%%%%%%%%%%%%%%%%%%
%%%%%%%%%%%%%%%%%%%%%%%%%%%%%%%%%%%%%%%%%%%%%%%%%%%%%%%%%%%%%%%%%%%%%%%%%%%%%%%

We assume in the following that the size of the interacting atoms or
molecules forming the dilute gas 
is much smaller than all other involved length scales 
such as the $s$-wave scattering length~$a_s$; typical energy
scales associated with scattering processes are thus assumed to be  
well below the internal energy differences.
In this limit, we may adopt the point-particle approximation and 
the interacting objects can be treated as structureless entities  
described by the field operator~$\hat\Psi$. 
Setting Planck's constant~$\hbar$ and the mass~$m$ of the atoms or
molecules equal to unity, the Heisenberg equation of motion for the
field operator reads  
\bea
\label{Heisenberg}
i\frac{\partial\hat\Psi}{\partial t}
&=&
\left(-\frac{\na^2}2 +V_{\rm ext}(\f{r})\right)\hat\Psi (\f{r})
\nn
&&+
\int d^3r'\,\hat\Psi^\dagger(\f{r}')V_{\rm int}(\f{r}-\f{r}')
\hat\Psi(\f{r}')\hat\Psi (\f{r})
\,,
\ea
with ${V_{\rm ext}(\f{r})}$ denoting the external one-particle
potential of the trap. 
The finite-range two-particle interaction 
potential~${V_{\rm int}(\f{r}-\f{r}')}$ is assumed to be sufficiently
``weak'' in the sense that the Born approximation applies, e.g., 
${V_{\rm int}(\f{r})}$ is supposed to be integrable 
(i.e., to decay more rapidly with distance than $1/r^3$) such that its Fourier 
transform~${\widetilde V_{\rm int}(\kf)}$ is a well-defined function
of~$\kf$.  
For large wavelengths, the finite-range two-particle interaction
potential~${V_{\rm int}(\f{r}-\f{r}')}$ yields an effective coupling
constant~$g$ determined by the $s$-wave scattering length~$a_s$
(in three spatial dimensions) 
\bea
\label{short-range}
g=4\pi a_s=\int d^3r\,V_{\rm int}(\f{r})=\widetilde V_{\rm int}(\kf=0)
\,.
\ea
The Bogoliubov mean-field approximation \cite{Bogoliubov} is based on
the assumption that the fluctuations~$\hat\chi$, describing
single-particle excitations above the ground state are small 
(compared to the mean field) if the number~$N$ of particles is large. 
In the following, we present a formal derivation of the
mean-field expansion for the system described in
Eq.\,(\ref{Heisenberg}) in the large-$N$~limit, i.e., an expansion of 
$\hat\Psi$ into powers of~$N$. 
We shall employ a particle-number-conserving mean-field
ansatz (satisfying ${\langle\hat\Psi\rangle=0}$) \cite{Particle,castin} 
\bea
\label{mean-field}
\hat\Psi=\left(\psi_0+\hat\chi+\hat\zeta\right)\frac{\hat A}{\sqrt{\hat N}}
\,.
\ea
Here, the {\em order parameter} is ${\psi_0}$. 
The single-particle excitations are described by $\hat\chi$, where 
{\em single-particle} here means that the Fourier components of 
  $\hat\chi$ 
are linear superpositions of annihilation and creation operators
of quasiparticles 
$\hat a_{\bm k}$ and $\hat a^\dagger_{\bm k}$, cf. Eq.\,(\ref{creation}). 
The higher-order corrections $\hat\zeta$ are due to multi-particle
excitations and correlations. Here ${\hat N=\hat A^\dagger\hat A}$
counts the total number of particles. 
%###

The {\em dilute-gas limit} may formally be defined by placing a very
large number $N\uparrow\infty$ of identical bosons into a finite trap
with $gN$ remaining constant \cite{castin,1/N}.  
The diluteness parameter of the gas then 
scales as $(g^3\varrho)^{1/2}\propto 1/N$, i.e., the gas becomes
rapidly more dilute with increasing particle number in this limit.  
This limit should be compared and contrasted with the
usual thermodynamic limit, in which the density and particle interaction 
remains constant, while the size of the (trapped) system increases
with $N\rightarrow \infty$, adjusting the trapping potential 
correspondingly (in $D$ spatial dimensions, the thermodynamic limit
corresponds to keeping $N\omega^D$  constant for $N\rightarrow \infty$, where
$\omega$ is the geometric mean of the trapping frequencies \cite{BECreview}). 
In the presently used  dilute-gas limit, on the other hand, 
the trapping potential remains constant, but the interaction 
and the density change.
The advantage of the limit $gN$ constant is that in this limit we have 
a well-defined prescription for checking the mean-field approximation, 
keeping one power of $g$ for each factor of $N$, cf.~\cite{Lee}. 

Given the dilute-gas limit, there arises a question which is central
to the applicability of the mean-field expansion:
Do the supposedly small multi-particle corrections~$\hat\zeta$
actually become small in this limit for a given behavior of the
interparticle potential ${V_{\rm int}(\f{r}-\f{r}')}$? 
In order to tackle this question, let us consider the following
scenario: 
Initially, the interaction is completely switched off,  
${V_{\rm int}(\f{r}-\f{r}')=0}$, and all $N$ bosons occupy the same
single-particle state (at zero temperature), described by the
single-particle wave-function ${\psi_0}/\sqrt{N}$.
In this case, Eq.\,(\ref{mean-field}) is trivially satisfied with 
${\psi_0=\ord(\sqrt{N})}$, ${\hat\chi=\ord(N^0)}$, $\hat\zeta=0$, and
${\hat A=\hat a_0\hat n_0^{-1/2}\hat N^{1/2}}$,  
where ${\hat n_0=\hat a_0^\dagger\hat a_0}$ corresponds to the
macroscopically occupied state,
% ###
cf.~\cite{Particle}. 
If we now switch on the interaction slowly enough, the system stays in
its ground state due to the adiabatic theorem and we may follow the
evolution of the field operator~$\hat\Psi$ and hence $\psi_0$,
$\hat\chi$, and $\hat\zeta$ (with $\hat A$ remaining constant, i.e., 
the evolution at given 
particle number) via insertion 
of Eq.\,(\ref{mean-field}) into~(\ref{Heisenberg}).

Assuming that the corrections~$\hat\zeta$ indeed become small for 
$N\uparrow\infty$, the leading order ($\sqrt{N}$) yields, 
in the dilute-gas limit,  
the Gross-Pitaevski\v\i\/ (GP) equation \cite{GP} for the order
parameter~$\psi_0$
\bea
\label{GP}
i\frac{\partial\psi_0}{\partial t}
&=&
\left(-\frac{\na^2}2 +V_{\rm ext}+g|\psi_0|^2\right)\psi_0
\,.
\ea
The next-to-leading order (${N^0}$) terms govern the evolution of 
the fluctuations $\hat\chi$ via the nonlocal analogue of the
Bogoliubov-de Gennes equations \cite{Bogoliubov,deGennes}
\bea
\label{BdG}
i\frac{\partial\hat\chi}{\partial t}
&=&
\int d^3r'
V_{\rm int}(\f{r}-\f{r}')\left[
|\psi_0|^2\hat\chi(\f{r}')+\psi_0^2\hat\chi^\dagger(\f{r}')
\right]\nn
&&+
\left(
-\frac{\na^2}2+V_{\rm ext} 
+g|\psi_0|^2\right)\hat\chi 
\,.
\ea
% ###
The remaining equation of motion for $\hat\zeta$ takes for a general
nonlocal potential the rather complicated form  
\bea
\label{zeta}
i\frac{\partial\hat\zeta}{\partial t}
&=&
\left(
-\frac{\na^2}{2}+V_{\rm ext}+|\psi_0|^2 
\left[g+\widehat V_{\rm int}*\right]
\right)\hat\zeta+\psi_0^2\widehat V_{\rm int}*\hat\zeta^\dagger
\nn
&&+
\psi_0^*\hat\chi\widehat V_{\rm int}*\hat\chi+
\psi_0(\widehat V_{\rm int}*\hat\chi^\dagger)\hat\chi+
\psi_0\widehat V_{\rm int}*(\hat\chi^\dagger\hat\chi)
\nn
&&+
\left[\widehat V_{\rm int}*(\hat\chi^\dagger\hat\chi)\right]\hat\chi
+\ord\left(\frac{\hat\zeta}{\sqrt{N}}\right)
\,,
\ea
where $\widehat V_{\rm int}*$ is an abbreviation for the convolution with
the quantity right from the ``$*$'' as in Eq.\,(\ref{BdG}), and 
${\ord(\zeta/\sqrt{N})}$ denote formally sub-leading terms containing 
$\hat\zeta$ or $\hat\zeta^\dagger$ or both.  
Observe that the commutation relations 
${[\hat\chi,\hat A{\hat N}^{-1/2}]=[\hat\zeta,\hat A{\hat N}^{-1/2}]=0}$ 
valid initially are preserved under the evolution given by the
equations above, i.e., the excitations $\hat\chi$ and $\hat\zeta$ are
always particle-number-conserving, cf.~\cite{Particle,castin}. 

If we start with free particles, and switch on the interaction
by following the evolution in Eqs.~(\ref{GP}), (\ref{BdG}), and
(\ref{zeta}), we observe a mixing of $\hat\chi$ and $\hat\chi^\dagger$
due to Eq.\,(\ref{BdG}), and $\hat\zeta$ does not vanish anymore due to the
four source terms 
${\hat\chi\widehat V_{\rm int}*\hat\chi}$, 
${(\widehat V_{\rm int}*\hat\chi^\dagger)\hat\chi}$, 
${\widehat V_{\rm int}*(\hat\chi^\dagger\hat\chi)}$, and
${[\widehat V_{\rm int}*(\hat\chi^\dagger\hat\chi)]\hat\chi}$. 
The viability of mean-field theory depends on the scaling of the
induced terms in $\hat\zeta$ with particle number, i.e., if and to
which extent they decrease for $N\uparrow\infty$. 
Naive power counting would imply ${\hat\zeta=\ord(1/\sqrt{N})}$, but
the nonlinearity due to the product of two or more field operators 
(such as in the above source terms) and the associated mode sum(s) may
compensate the smallness of ${1/\sqrt{N}}$.
As an example, let us consider the expectation values of the four
source terms.
The last term ${[\widehat V_{\rm int}*(\hat\chi^\dagger\hat\chi)]\hat\chi}$
is both sub-leading and odd so that its expectation value vanishes.  
However, the expectation values of the remaining terms do not vanish, 
and affect ${\langle\hat\zeta (N)\rangle }$. 

For the sake of notational  
convenience, we consider in what follows a homogeneous condensate 
${\psi_0=\sqrt{\varrho_0}=\rm const}$ in a constant ``trapping''
potential ${V_{\rm ext}=-g\varrho_0}$, so that 
%the chemical potential vanishes, ${\mu=0}$ and 
the two last terms in the 
round brackets on the right-hand side of the 
Bogoliubov-de Gennes equations (\ref{BdG}) cancel.  
A normal-mode expansion for the fluctuation operators then yields 
%for the Bogoliubov-de Gennes equations (\ref{BdG}) 
in Fourier space
\bea
\label{BdG-k}
i\frac{\partial\hat\chi_\kf}{\partial t}
=
\left(
\frac{\kf^2}{2}+\varrho_0\widetilde V_{\rm int}(\kf)
\right)\hat\chi_\kf
+\varrho_0\widetilde V_{\rm int}(\kf)\hat\chi_\kf^\dagger
\,,
\ea
so that the annihilation operators $\hat\chi_\kf$ of the original bosons
% ### bosonic excitations 
have the following Bogoliubov transformation form in terms of the  
quasiparticle operators $\hat a_{\bm k}, \hat a^\dagger_{\bm k}$:
\bea
\label{creation}
\hat\chi_{\boldsymbol k}
=
\sqrt{\frac{{\boldsymbol k}^2}{2\omega_{\boldsymbol k}}}
\left[
\left(\frac12-\frac{\omega_{\boldsymbol k}}{{\boldsymbol k}^2}\right)
\hat a_{\boldsymbol k}^\dagger
+
\left(\frac12+\frac{\omega_{\boldsymbol k}}{{\boldsymbol k}^2}\right)
\hat a_{\boldsymbol k}
\right]
\,.
\ea
The generalized Bogoliubov dispersion relation reads 
\bea
\label{Bogoliubov}
\omega_{\boldsymbol k}^2=
{\boldsymbol k}^2\varrho_0\widetilde V_{\rm int}(\kf)
+\frac{{\boldsymbol k}^4}{4}
\,.
\ea
For wavenumbers $\kf^2\xi^2\gg1$ much larger than the inverse of the
healing length ${\xi=1/\sqrt{g\varrho_0}}$, we have 
${\hat\chi_{\boldsymbol k}\sim\hat a_{\boldsymbol k}-
\hat a_{\boldsymbol k}^\dagger
\varrho_0\widetilde V_{\rm int}(\kf)/{\boldsymbol k}^2}$, 
i.e., the quasiparticles become asymptotically equal to the original
bosons.  
However, this $1/{\boldsymbol k}^2$ decrease alone is not sufficient
for rendering the expectation value of 
${\langle\hat\chi\widehat V_{\rm int}*\hat\chi\rangle}$
finite (except in one spatial dimension).

A smooth (and integrable) interaction potential $V_{\rm int}(\f{r})$
implies a faster-than-polynomial decrease 
${\widetilde V_{\rm int}(\kf^2 \gg k_{\rm cut}^2)=0}$
for large wavenumbers ${\kf^2 \gg k_{\rm cut}^2}$.
The {\em cut-off wavenumber} $k_{\rm cut}$, then,  is determined by the first
significant deviations of ${\widetilde V_{\rm int}}$ from its
long-wavelength behavior  
${\widetilde V_{\rm int}(\kf^2 \ll k_{\rm cut}^2)=g}$
and is assumed to be much larger than the inverse of the healing length. 
This faster-than-polynomial decrease renders all relevant expectation
values finite. 
For example, the expectation value of the so-called ``anomalous'' term
reads  
\bea
\label{anomal}
\langle\hat\chi\widehat V_{\rm int}*\hat\chi\rangle
=
-\frac{\psi_0^2}{2\pi^2}\int dk\,
\frac{k^2\widetilde V_{\rm int}^2(k)}{2\omega_k}
\,,
\ea
where we have assumed a spherically symmetric potential 
$V_{\rm int}(r)$ and hence ${\widetilde V_{\rm int}}=
{\widetilde V_{\rm int}(k)}$.  
The expectation value is taken in the ground state of the
quasiparticles (remember the adiabatic switching process) 
which is annihilated by $\hat a_{\boldsymbol k}$ 
(but not by $\hat\chi_{\boldsymbol k}$, of course).
For a sufficiently regular ${\widetilde V_{\rm int}(k)}$, an expansion 
into powers of $1/(\xi k_{\rm cut})$ yields 
\bea
\label{anomal-expand}
\langle\hat\chi\widehat V_{\rm int}*\hat\chi\rangle
=
-\frac{\psi_0^2}{2\pi^2}\int dk\,\widetilde V_{\rm int}^2
+
\frac{g^2\psi_0^2}{\xi\pi^2}
+\ord\left(\frac{g^3\varrho_0^2}{k_{\rm cut}}\right).
\ea
The $1/(\xi k_{\rm cut})$-corrections depend on the explicit form of 
$V_{\rm int}(r)$ and hence ${\widetilde V_{\rm int}(k)}$, e.g., for
${\widetilde V_{\rm int}(k)=g\Theta(k_{\rm cut}-k)}$, 
the integral in Eq.\,(\ref{anomal}) yields 
$g^2(\sqrt{\xi^2k_{\rm cut}^2+4}-2)/\xi$.

In contrast to the so-called ``anomalous'' term 
${\langle\hat\chi\widehat V_{\rm int}*\hat\chi\rangle}$, 
the expectation value of the quantum depletion terms 
${\widehat V_{\rm int}*\langle\hat\chi^\dagger\hat\chi\rangle}$
and 
${\langle(\widehat V_{\rm int}*\hat\chi^\dagger)\hat\chi\rangle}$
occurring as source terms in Eq.\,(\ref{zeta})
do not have a contribution linear in $k_{\rm cut}$.
Hence their (for ${\xi k_{\rm cut}\gg1}$) dominant terms are
independent of $k_{\rm cut}$ and both give 
${g\langle\hat\chi^\dagger\hat\chi\rangle\approx
\sqrt{g^5\varrho_0^3}/(3\pi^2)}$.
The higher-order $1/(\xi k_{\rm cut})$-corrections, however, again
depend on the shape of ${\widetilde V_{\rm int}({\bm k})}$ and can be
calculated analogously. 

%%%%%%%%%%%%%%%%%%%%%%%%%%%%%%%%%%%%%%%%%%%%%%%%%%%%%%%%%%%%%%%%%%%%%%%%%%%%%%%
%%%%%%%%%%%%%%%%%%%%%%%%%%%%%%%%%%%%%%%%%%%%%%%%%%%%%%%%%%%%%%%%%%%%%%%%%%%%%%%
\section{Consistency of mean-field expansion} 
%%%%%%%%%%%%%%%%%%%%%%%%%%%%%%%%%%%%%%%%%%%%%%%%%%%%%%%%%%%%%%%%%%%%%%%%%%%%%%%
%%%%%%%%%%%%%%%%%%%%%%%%%%%%%%%%%%%%%%%%%%%%%%%%%%%%%%%%%%%%%%%%%%%%%%%%%%%%%%%

To explicitly address our principal question of whether the higher-order
corrections~$\hat\zeta$ in the mean-field expansion~(\ref{mean-field})
indeed become small in the large-$N$~limit (in which one would  
naively expect mean-field to become more and more accurate), we now
discuss three 
concrete examples for the formal scaling behavior of the interaction 
$V_{\rm int}$ with the number of particles $N$. 
To this end, the particle-number-dependent potential $V_{\rm int}^{(N)}$ 
is always chosen such that it reproduces the $g\propto 1/N$
dilute-gas limit prescription for the behavior of the coupling
constant in the large-$N$~limit.

In the first example, we assume $k_{\rm cut}$ to be (formally)
independent of $N$  
\bea
\label{independent}
V_{\rm int}^{(N)}(\f{r})=\frac1N\,V_{\rm int}^{(1)}(\f{r})
\,.
\ea
Since $k_{\rm cut}$ remains constant for $N\uparrow\infty$, the mode
sums due to the nonlinear terms in Eq.\,(\ref{zeta}) cannot compensate
the smallness of the pre-factors such as $1/\sqrt{N}$ and thus we
obtain  
\bea
\label{1/2}
\hat\zeta=\ord\left(\frac{1}{\sqrt{N}}\right)
\,.\label{first}
\ea
However, this decrease is not sufficient yet for ensuring
the usual split of the total density 
\bea
\label{density}
\varrho
=
\left\langle\hat\Psi^\dagger\hat\Psi\right\rangle
=
|\psi_0|^2+\left\langle\hat\chi^\dagger\hat\chi\right\rangle+
\ord\left(\frac{1}{\sqrt{N}}\right)
\,,
\ea
since ${\psi_0\langle\hat\zeta\rangle}$ could be of the same order as 
$\langle\hat\chi^\dagger\hat\chi\rangle$.
For the above split, we need $\langle\hat\zeta\rangle=\ord(1/N)$
instead of $\ord(1/\sqrt{N})$, which requires absorbing the
expectation values of the source terms into a {\em modified} GP equation   
\bea
\label{GP-mod}
i\frac{\partial\psi_0}{\partial t} 
&=&
\left(
-\frac{\na^2}2 +V_{\rm ext}+g|\psi_0|^2
+g\left\langle\hat\chi^\dagger\hat\chi\right\rangle
\right)\psi_0
\nn
&&+
\langle\hat\chi\widehat V_{\rm int}*\hat\chi\rangle\psi_0^*
+
\langle(\widehat V_{\rm int}*\hat\chi^\dagger)\hat\chi\rangle\psi_0
\,.
\ea
Inserting Eq.\,(\ref{anomal-expand}), we see that the dominant 
(for  $\xi k_{\rm cut}\gg1$) and density-independent
fluctuation contribution 
$\langle\hat\chi\widehat V_{\rm int}*\hat\chi\rangle$
proportional to $k_{\rm cut}$ can be absorbed by a
renormalization of the coupling constant
\bea
\label{renormalization}
g_{\rm ren}
&=&
g-\frac{1}{2\pi^2}\int dk\,\widetilde V_{\rm int}^2
\nn
&=&
g+\int d^3r\;V_{\rm int}\na^{-2}V_{\rm int}
\,,
\ea
where the second line is due to Parseval's theorem,
using the formal inverse of the Laplace operator $\na^2$. 
Since $-\na^{-2}/2$ and $V_{\rm int}/2$ are the kernels of the 
Schr\"odinger Hamiltonian~$\hat H_0$ and the interaction
Hamiltonian~$\hat H_1$, respectively, this just corresponds to the
usual one-loop renormalization
%$\langle\hat H_1\rangle-\langle\hat H_1\hat H_0^{-1}\hat H_1\rangle$
of the interaction potential.
%###
Note that in three spatial dimensions, sub-dominant contributions 
like $\langle\hat\chi^\dagger\hat\chi\rangle$ cannot be absorbed by
such a renormalization procedure.  

The physical significance of the renormalized coupling $g_{\rm ren}$
can be demonstrated further by calculating the 
$k_{\rm cut}$-corrections to the total energy of a homogeneous gas 
\bea
\label{energy}
E
=
\langle\hat H\rangle
\approx
NV_{\rm ext}+\frac{N}{2}g\varrho-
\frac{N}{4\pi^2}\varrho\int dk\,\widetilde V_{\rm int}^2
\,,
\ea
which gives $E=NV_{\rm ext}+Ng_{\rm ren}\varrho/2$ plus
higher-order corrections (cf.\,\cite{Lee}). 

A potential complication which arises from the 
prescription~(\ref{independent}) is that the range of the two-particle
interaction potential in Eq.\,(\ref{independent}) will exceed the
inter-particle distance~$d_i$ for large~$N$ since ${\varrho_0\sim N}$
and thus ${d_i\sim1/N^{1/3}}$ (in three spatial dimensions). 
If the two-particle interaction potential is mainly caused by direct
collisions, this might lead to a conflict with the point-particle
approximation used in writing down the starting point of the analysis,
Eq.\,(\ref{Heisenberg}). 
Therefore, as a second example, consider the support of 
${V_{\rm int}(\f{r})}$ to be decreasing in proportion to the
inter-particle distance $d_i$,   
\bea
\label{scaling}
V_{\rm int}^{(N)}(\f{r})=V_{\rm int}^{(1)}(N^{1/3}\f{r})
\,.
\ea
As a consequence, the cut-off scales as ${k_{\rm cut}\sim N^{1/3}}$,
and thus 
${\langle\hat\chi\widehat V_{\rm int}*\hat\chi\rangle\sim N^{-2/3}}$, 
%###
which ensures that the remaining $\hat\zeta$ corrections are still
small, though decreasing with a smaller power in $N$ than in the first
example Eq.\,(\ref{first}):  
\bea
\label{1/6}
\hat\zeta=\ord\left(\frac{1}{N^{1/6}}\right)
\,.
\ea
The above scaling with particle number is due to the fact that 
for each additional operator in Eq.\,(\ref{zeta}), which might
contribute a factor ${k_{\rm cut}\sim N^{1/3}}$ after the mode
summation, there is a pre-factor of order $1/\sqrt{N}$. 
In this case, the split in Eq.\,(\ref{density}) is still possible
provided the modification of the GP equation~(\ref{GP-mod})
is employed, but the estimate of the accuracy is now 
the rather slow decrease of $\ord(1/N^{1/6})$
instead of $\ord(1/N^{1/2})$.

Finally, as a third example, we investigate a scaling employed by 
Lieb et al.~\cite{1/N,Lieb}, used in a proof 
of the asymptotic exactness of the Gross-Pitaevski\v\i\/
energy functional in three spatial dimensions, and in a analysis of 
one-dimensional systems of bosons in 3D traps, respectively. 
This scaling reads  
\bea
\label{Lieb}
V_{\rm int}^{(N)}(\f{r})=N^2V_{\rm int}^{(1)}(N\f{r})
\,, 
\ea
and implies that the cut-off increases linearly with particle
number, ${k_{\rm cut}\sim N}$, and thus 
${\langle\hat\chi\widehat V_{\rm int}*\hat\chi\rangle\sim N^0}$,
%### 
so that the anomalous term becomes of the same order as other
terms in the GP equation, for example as large as the mean-field
interaction term~$g|\psi_0|^2$. 
The ``correction'' operator then behaves as
\bea
\label{1/1}
\hat\zeta=\ord\left(\sqrt{N}\right)
\,,\label{third}
\ea
and we have no true control over the corrections in the mean-field
expansion which ought to be negligible. 
The mean-field approximation can only be consistent (if at all) with a 
modified GP equation~(\ref{GP-mod}) inducing a renormalization of the
coupling according to Eq.\,(\ref{renormalization}). 
However, even given these modifications, the applicability of the
mean-field expansion is not obvious since higher-order operator
products can yield $\ord(N)$ contributions after the
$k$-summation/integration and thus the hierarchy
$\hat\zeta\ll\hat\chi$ is not evident. 
Similarly, the first-order correction in Eq.\,(\ref{renormalization}) 
is comparable to the zeroth order (i.e., of the same order in~$N$),
which hints at the fact that all orders must be taken into account in
a suitable way.  

%###
It is illuminating to compare the employed dilute-gas limit 
($N\uparrow\infty$ particles in a fixed volume with $gN$ remaining constant)
with the thermodynamic limit 
($N\uparrow\infty$ particles in an increasing volume $\cal V$ with $g$
and $N/\cal V$ remaining constant):
Translation of the scaling in Eq.\,(\ref{independent}) to the
thermodynamic limit yields an interaction potential whose range
increases proportional to the system size $\cal V$ whereas its
strength decreases accordingly.
It is not very surprising that the mean-field expansion is very good
in this case. 
The analogue of Eq.\,(\ref{scaling}) in the thermodynamic limit is an
interaction potential with constant strength and range (where the
applicability of the mean-field expansion is less obvious).
Finally, Eq.\,(\ref{Lieb}) corresponds to a potential with decreasing
range and increasing strength (in the thermodynamic limit).
Again, is not very surprising that such a scaling might generate
difficulties in the ordinary mean-field expansion and requires taking
into accout all orders in a suitable way.
%###

%%%%%%%%%%%%%%%%%%%%%%%%%%%%%%%%%%%%%%%%%%%%%%%%%%%%%%%%%%%%%%%%%%%%%%%%%%%%%%%
%%%%%%%%%%%%%%%%%%%%%%%%%%%%%%%%%%%%%%%%%%%%%%%%%%%%%%%%%%%%%%%%%%%%%%%%%%%%%%%
\section{Dipole-dipole interaction} 
%%%%%%%%%%%%%%%%%%%%%%%%%%%%%%%%%%%%%%%%%%%%%%%%%%%%%%%%%%%%%%%%%%%%%%%%%%%%%%%
%%%%%%%%%%%%%%%%%%%%%%%%%%%%%%%%%%%%%%%%%%%%%%%%%%%%%%%%%%%%%%%%%%%%%%%%%%%%%%%

As an example for interactions with a finite range, let us consider
a dipole-dipole force~${\propto g_d}$ in addition to the usual 
contact repulsion~${\propto g}$. 
Dipole-dipole interactions between atoms \cite{YiYou1}
can either be induced by an external electric field or be due to an
intrinsic magnetic dipole moment. 
E.g., for magnetic dipoles, ${g_d=\mu_0 d_m^2/3}$; Bose-Einstein
condensation of chromium, which has a ground state moment of 
$d_m=6\mu_B$, has been achieved recently \cite{Griesmaier}.  

If the dipole moments of all atoms/molecules are aligned along the
$z$-axis, the dipole-dipole interaction potential reads 
(in three spatial dimensions)
\begin{equation} 
V_{dd} ({\bm r}) = \frac{3g_d}{4\pi} 
\frac{1-3z^2/|{\bm r}|^2}{|{\bm r}|^3} 
\,. 
\label{Vdd}
\end{equation}  
In addition to this long-range interaction, the particles are subject
to a short-range repulsion whose impact can be represented by a
contact potential $\propto\delta^3({\bm r}-{\bm r}')$. 
Consequently, the Fourier transformed potential reads for intermediate
momenta 
\begin{equation} 
{\widetilde V}_{\rm int}({\bm k})=g+g_d\left(\frac{3k_z^2}{{\bm k}^2}-1\right).
\end{equation}
The (ideal) dipole-dipole interaction potential
behaves as~$1/r^3$ which is not integrable and, strictly speaking, is   
therefore just at the limit 
of applicability of our analysis -- for example, 
${\widetilde V}_{\rm int}$ is not well-defined at $\f{k}=0$, which
complicates the introduction of an effective coupling~$g$
(finite-size effects etc.)
Nevertheless, we may regard the nonintegrable dipole-dipole
interaction potential as a limit of integrable potentials and
calculate the corresponding corrections beyond the $s$-wave scattering  
approximation.
For example, the quantum depletion is modified from 
the pure contact case via 
\bea
\label{dipole}
\langle\hat\chi^\dagger\hat\chi\rangle
&=& 
\frac{\varrho_0^{3/2}}{24\pi^2}\,\sqrt{2g_d + g}\,(g_d+5g)
\nn
&&-
\frac{\varrho_0^{3/2}}{16\pi^2}\,\frac{(g-g_d)^2}{\sqrt{3g_d}}\, 
\ln\left[
\frac{g-g_d}{(\sqrt{3g_d}+\sqrt{2g_d+g}\,)^2}
\right]
\nn
&=& 
\frac{(g\varrho_0)^{3/2}}{3\pi^2}
\left[
1+\frac{3}{10}\,\frac{g_d^2}{g^2}+\ord\left(\frac{g_d^3}{g^3}\right)
\right]
\,.
\ea
This expression is valid for $g \geq g_d$ only, since
otherwise the excitation spectrum (for a homogeneous condensate) and
hence also ${\langle\hat\chi^\dagger\hat\chi\rangle}$
contain imaginary parts indicating an instability, cf.~\cite{Goral}.  
%
% 
%Based on the derivation presented in this article, f
Further expectation values such as the total energy 
${E=\langle\hat H\rangle}$ can be calculated analogously \cite{Fischer}. 

%%%%%%%%%%%%%%%%%%%%%%%%%%%%%%%%%%%%%%%%%%%%%%%%%%%%%%%%%%%%%%%%%%%%%%%%%%%%%%%
%%%%%%%%%%%%%%%%%%%%%%%%%%%%%%%%%%%%%%%%%%%%%%%%%%%%%%%%%%%%%%%%%%%%%%%%%%%%%%%
\section{Conclusion}
%%%%%%%%%%%%%%%%%%%%%%%%%%%%%%%%%%%%%%%%%%%%%%%%%%%%%%%%%%%%%%%%%%%%%%%%%%%%%%%
%%%%%%%%%%%%%%%%%%%%%%%%%%%%%%%%%%%%%%%%%%%%%%%%%%%%%%%%%%%%%%%%%%%%%%%%%%%%%%%

Based on the point-particle approximation of Eq.~(\ref{Heisenberg}),
we derived the mean-field expansion Eq.~(\ref{mean-field}) for
dilute Bose-Einstein condensates with arbitrary weak finite-range
two-particle interaction potentials, obeying suitable scaling behavior
in the large-$N$ limit. It turns out that, although the 
gas rapidly becomes ever more dilute in the large $N$ limit (the gas
parameter $(g^3\varrho)^{1/2}\propto 1/N$), the validity of the mean-field
approximation strongly depends on the detailed scaling behavior of the 
particle interaction potential. Therefore, care needs to be exercised 
in applying the mean-field approximation -- if one does not  
take the detailed structure of the bare interaction 
potential into account, one possibly encounters 
inconsistencies with the basic assumptions 
the mean-field expansion is built on.

Apart from exploring limits of the mean-field approximation, our 
derivation facilitates the calculation of the impact of microscopic
details of the interaction beyond the $s$-wave scattering
approximation \cite{Geltman}. 
Another advantage of the presented approach is the natural 
emergence and {\em ab initio} derivation of the cut-off~$k_{\rm cut}$ 
as a microscopic property of the interaction potential -- instead of a 
cut-off $k_{\rm cut}$ introduced {\em ad hoc} for the regularization
of pseudo-potentials $\propto {\delta^3(\f{r}-\f{r}')}$, see, e.g.,
\cite{Braaten}. 
However, one should bear in mind that the presented method requires
weak potentials -- if nonperturbative effects such as bound states or
total reflection at a finite radius become important, the direct
mean-field ansatz~(\ref{mean-field}) cannot be applied in this way.
Instead of homogeneous plane waves, one has to start from
atomic/molecular eigenfunctions in this case and the
ansatz~(\ref{mean-field}) can only work as a low-energy effective
description.  
Nevertheless, if it is possible to divide a nonperturbative 
${V_{\rm int}(\f{r}-\f{r}')}$ into a strong ultra-short-range part and
a comparably weak remaining contribution which is more spread out, the
presented analysis should be applicable to the latter part, provided the 
ultra-short-range part can be replaced in an adequate manner by an integrable 
(pseudo-)potential. An example for the application of our analysis to the 
longer-range contribution we gave in the last Section,
using a dipole-dipole interaction potential.

%%%%%%%%%%%%%%%%%%%%%%%%%%%%%%%%%%%%%%%%%%%%%%%%%%%%%%%%%%%%%%%%%%%%%%%%%%%%%%%
%%%%%%%%%%%%%%%%%%%%%%%%%%%%%%%%%%%%%%%%%%%%%%%%%%%%%%%%%%%%%%%%%%%%%%%%%%%%%%%
\acknowledgments
%%%%%%%%%%%%%%%%%%%%%%%%%%%%%%%%%%%%%%%%%%%%%%%%%%%%%%%%%%%%%%%%%%%%%%%%%%%%%%%
%%%%%%%%%%%%%%%%%%%%%%%%%%%%%%%%%%%%%%%%%%%%%%%%%%%%%%%%%%%%%%%%%%%%%%%%%%%%%%%
R.\,S., M.\,U., and Y.\,X. gratefully acknowledge financial support by
the Emmy Noether Programme of the German Research Foundation (DFG)
under grant No.~SCHU~1557/1-1.
R.\,S.~and U.\,R.\,F.~acknowledge support by the COSLAB programme of the  
ESF.   

%\newpage

%%%%%%%%%%%%%%%%%%%%%%%%%%%%%%%%%%%%%%%%%%%%%%%%%%%%%%%%%%%%%%%%%%%%%%%%%%%%%%%
%%%%%%%%%%%%%%%%%%%%%%%%%%%%%%%%%%%%%%%%%%%%%%%%%%%%%%%%%%%%%%%%%%%%%%%%%%%%%%%
%%%%%%%%%%%%%%%%%%%%%%%%%%%%%%%%%%%%%%%%%%%%%%%%%%%%%%%%%%%%%%%%%%%%%%%%%%%%%%%
%%%%%%%%%%%%%%%%%%%%%%%%%%%%%%%%%%%%%%%%%%%%%%%%%%%%%%%%%%%%%%%%%%%%%%%%%%%%%%%
%%%%%%%%%%%%%%%%%%%%%%%%%%%%%%%%%%%%%%%%%%%%%%%%%%%%%%%%%%%%%%%%%%%%%%%%%%%%%%%
%%%%%%%%%%%%%%%%%%%%%%%%%%%%%%%%%%%%%%%%%%%%%%%%%%%%%%%%%%%%%%%%%%%%%%%%%%%%%%%
%%%%%%%%%%%%%%%%%%%%%%%%%%%%%%%%%%%%%%%%%%%%%%%%%%%%%%%%%%%%%%%%%%%%%%%%%%%%%%%
%%%%%%%%%%%%%%%%%%%%%%%%%%%%%%%%%%%%%%%%%%%%%%%%%%%%%%%%%%%%%%%%%%%%%%%%%%%%%%%
%%%%%%%%%%%%%%%%%%%%%%%%%%%%%%%%%%%%%%%%%%%%%%%%%%%%%%%%%%%%%%%%%%%%%%%%%%%%%%%
%%%%%%%%%%%%%%%%%%%%%%%%%%%%%%%%%%%%%%%%%%%%%%%%%%%%%%%%%%%%%%%%%%%%%%%%%%%%%%%
%%%%%%%%%%%%%%%%%%%%%%%%%%%%%%%%%%%%%%%%%%%%%%%%%%%%%%%%%%%%%%%%%%%%%%%%%%%%%%%
%%%%%%%%%%%%%%%%%%%%%%%%%%%%%%%%%%%%%%%%%%%%%%%%%%%%%%%%%%%%%%%%%%%%%%%%%%%%%%%
%%%%%%%%%%%%%%%%%%%%%%%%%%%%%%%%%%%%%%%%%%%%%%%%%%%%%%%%%%%%%%%%%%%%%%%%%%%%%%%
%%%%%%%%%%%%%%%%%%%%%%%%%%%%%%%%%%%%%%%%%%%%%%%%%%%%%%%%%%%%%%%%%%%%%%%%%%%%%%%
%%%%%%%%%%%%%%%%%%%%%%%%%%%%%%%%%%%%%%%%%%%%%%%%%%%%%%%%%%%%%%%%%%%%%%%%%%%%%%%
%%%%%%%%%%%%%%%%%%%%%%%%%%%%%%%%%%%%%%%%%%%%%%%%%%%%%%%%%%%%%%%%%%%%%%%%%%%%%%%


\begin{thebibliography}{499}

\bibitem{Bogoliubov} 
N.\,N.~Bogoliubov, J.\ Phys.\ (USSR) {\bf 11}, 23 (1947).

\bibitem{Lee} 
%T.\,D.~Lee and C.\,N.~Yang, 
%Many-Body Problem in Quantum Mechanics and Quantum Statistical
%Mechanics 
%Phys.\ Rev.\ {\bf 105}, 1119 (1957);
%
T.\,D.~Lee, K.~Huang, and C.\,N.~Yang, 
%Eigenvalues and Eigenfunctions of a Bose System of Hard Spheres and
%Its Low-Temperature Properties 
%{\it ibid.}\ 
Phys. Rev. {\bf 106}, 1135 (1957). 

\bibitem{Beliaev} 
S.\,T.~Beliaev, Sov.\ Phys.\ JETP {\bf 7}, 289 (1958).
%APPLICATION OF THE METHODS OF QUANTUM FIELD THEORY TO A SYSTEM OF BOSONS

\bibitem{Andersen} 
J.\,O.~Andersen, 
Rev.\ Mod.\ Phys.\ {\bf 76}, 599 (2004); contains
a selection of references of more recent work.

\bibitem{BECreview} 
F.~Dalfovo, S.~Giorgini, L.~P.~Pitaevski\v{\i}, and 
S.~Stringari, 
%: {\em Theory of trapped Bose-condensed gases}, %cond-mat/9806038, 
Rev.\ Mod.\ Phys.\ {\bf 71}, 463 (1999).  

\bibitem{Griffin} 
A.~Griffin, Phys.\ Rev.\ B {\bf 53}, 9341 (1996). 

\bibitem{Rusch} 
M. Rusch, S.\,A. Morgan, 
D.\,A.\,W. Hutchinson, and K. Burnett, Phys. Rev. Lett. {\bf 85}, 4844 (2000); 
D.\,A.\,W. Hutchinson {\it et al.},  J. Phys. B %: At. Mol. Opt. Phys. 
{\bf 33}, 3825 (2000); 
S.\,A. Morgan, J. Phys. B %: At. Mol. Opt. Phys. 
{\bf 33}, 3847 (2000). 

\bibitem{Olshanii} 
M. Olshani\v\i\/ and L. Pricoupenko, 
Phys. Rev. Lett. {\bf 88}, 010402 (2002). 

\bibitem{GP} 
E.~P.~Gross,
Nuovo Cimento {\bf 20}, 454 (1961);
%{\em Hydrodynamics of a superfluid condensate},
J.\ Math.\ Phys.\ {\bf 4}, 195 (1963);
L.~P.~Pitaevski\v\i\/,
Zh.\ Eksp.\ Teor.\ Fiz.\ {\bf 40}, 646 (1961);
Sov.\ Phys.\ JETP {\bf 13}, 451 (1961).

\bibitem{deGennes} 
P.\,G.~de~Gennes,
{\em Superconductivity of Metals and Alloys}
(W.\,A. Benjamin, New York, 1966).

\bibitem{Particle} 
M.~Girardeau and R.~Arnowitt,
%{\em Theory of Many-Boson Systems: Pair Theory},
Phys.\ Rev.\ {\bf 113}, 755 (1959);
%
C.\,W.~Gardiner, 
%{\em Particle-number-conserving Bogoliubov method which demonstrates 
%the validity of the time-dependent Gross-Pitaevski\v\i\/ equation for a 
%highly condensed Bose gas},
Phys.\ Rev.\ A {\bf 56}, 1414 (1997);
%
M.\,D.~Girardeau, 
%{\em Comment on "Particle-number-conserving Bogoliubov method which 
%demonstrates the validity of the time-dependent Gross-Pitaevski\v\i\/
%equation  for a highly condensed Bose gas"},
{\it ibid.}\ {\bf 58}, 775 (1998). 

\bibitem{castin} 
Y.~Castin and R.~Dum,
%Low-temperature Bose-Einstein condensates in time-dependent traps:
%Beyond the U(1) symmetry-breaking approach 
Phys.\ Rev.\ A {\bf 57}, 3008 (1998).

\bibitem{1/N} 
E.\,H.~Lieb, R.~Seiringer, and J.~Yngvason, 
%{\em Bosons in a trap: A rigorous derivation of the Gross-Pitaevski\v\i\/
%energy functional},
Phys.\ Rev.\ A {\bf 61}, 043602 (2000).

\bibitem{Lieb} 
E.\,H.~Lieb, R.~Seiringer, and J.~Yngvason, 
% One-Dimensional Bosons in Three-Dimensional Traps
Phys.\ Rev.\ Lett.\ {\bf 91}, 150401 (2003). 

\bibitem{YiYou1} 
S.~Yi and L.~You, Phys.\ Rev.\ A {\bf 61}, 041604(R) (2000).

\bibitem{Griesmaier}
A. Griesmaier {\it et al.}, 
%J. Werner, S. Hensler, J. Stuhler, and T. Pfau, 
%cond-mat/0503044. 
Phys. Rev. Lett. {\bf 94}, 160401 (2005).

\bibitem{Goral} 
K.~G\'oral, K.~Rz\c a$\dot {\rm z}$ewski, and T.~Pfau, 
Phys.\ Rev.\ A {\bf 61}, 051601(R) (2000).

\bibitem{Fischer}
We note that, by reducing the three-dimensional 
geometry discussed here to a quasi-two-dimensional one, 
the bare dipole-dipole interaction is converted into a  
finite-range potential with regular
Fourier transform, where the range of the potential %in this case 
is given by the thickness of the system in the strongly confining direction.
The quantum depletion %(\ref{dipole}) 
then remains regular and real  
even for $g_d \gg g $, and the analysis of the present paper 
applies also in this region of parameter space, see 
U.~R.~Fischer,
%: {\em Stability and existence of quasi-two-dimensional Bose-Einstein
%  condensates with dominant dipole-dipole interactions}, 
{\sf preprint} cond-mat/0506368.  
%###

\bibitem{Geltman} 
The inadequacy of $a_s$ as the sole parameter to describe
Bose-Einstein condensates has also been discussed by 
S.~Geltman and A.~Bambini,  
%: {\em Triplet Scattering Lengths
%for Rubidium and their Role in Bose-Einstein Condensation}, 
Phys.\ Rev.\ Lett.\ {\bf 86}, 3276 (2001). 
% ###
They do however not address the actual {\em applicability} 
of the mean-field approach.

\bibitem{Braaten} 
E.~Braaten and A.~Nieto, 
Phys.\ Rev.\ B {\bf 56}, 14745 (1997).

\end{thebibliography}
\end{document}